\RequirePackage{lineno}
\documentclass[aps,prd,twocolumn,superscriptaddress,showpacs,preprintnumbers,amsmath,amssymb,floatfix]{revtex4-1}

\usepackage{subfigure}
\usepackage{graphicx} 
\usepackage{dcolumn}  
\usepackage{nccmath}  

\graphicspath{{ps}}

\usepackage{xcolor}
\begin{document}



\title{ \quad\\[1.0cm]Study of Electromagnetic Decays of Orbitally Excited $\Xi_c$ Baryons}


\noaffiliation
\affiliation{University of the Basque Country UPV/EHU, 48080 Bilbao}
\affiliation{University of Bonn, 53115 Bonn}
\affiliation{Brookhaven National Laboratory, Upton, New York 11973}
\affiliation{Budker Institute of Nuclear Physics SB RAS, Novosibirsk 630090}
\affiliation{Faculty of Mathematics and Physics, Charles University, 121 16 Prague}
\affiliation{Chonnam National University, Gwangju 61186}
\affiliation{University of Cincinnati, Cincinnati, Ohio 45221}
\affiliation{Deutsches Elektronen--Synchrotron, 22607 Hamburg}
\affiliation{University of Florida, Gainesville, Florida 32611}
\affiliation{Department of Physics, Fu Jen Catholic University, Taipei 24205}
\affiliation{Key Laboratory of Nuclear Physics and Ion-beam Application (MOE) and Institute of Modern Physics, Fudan University, Shanghai 200443}
\affiliation{Justus-Liebig-Universit\"at Gie\ss{}en, 35392 Gie\ss{}en}
\affiliation{Gifu University, Gifu 501-1193}
\affiliation{II. Physikalisches Institut, Georg-August-Universit\"at G\"ottingen, 37073 G\"ottingen}
\affiliation{SOKENDAI (The Graduate University for Advanced Studies), Hayama 240-0193}
\affiliation{Gyeongsang National University, Jinju 52828}
\affiliation{Department of Physics and Institute of Natural Sciences, Hanyang University, Seoul 04763}
\affiliation{University of Hawaii, Honolulu, Hawaii 96822}
\affiliation{High Energy Accelerator Research Organization (KEK), Tsukuba 305-0801}
\affiliation{J-PARC Branch, KEK Theory Center, High Energy Accelerator Research Organization (KEK), Tsukuba 305-0801}
\affiliation{Higher School of Economics (HSE), Moscow 101000}
\affiliation{Forschungszentrum J\"{u}lich, 52425 J\"{u}lich}
\affiliation{IKERBASQUE, Basque Foundation for Science, 48013 Bilbao}
\affiliation{Indian Institute of Science Education and Research Mohali, SAS Nagar, 140306}
\affiliation{Indian Institute of Technology Bhubaneswar, Satya Nagar 751007}
\affiliation{Indian Institute of Technology Guwahati, Assam 781039}
\affiliation{Indian Institute of Technology Hyderabad, Telangana 502285}
\affiliation{Indian Institute of Technology Madras, Chennai 600036}
\affiliation{Indiana University, Bloomington, Indiana 47408}
\affiliation{Institute of High Energy Physics, Chinese Academy of Sciences, Beijing 100049}
\affiliation{Institute of High Energy Physics, Vienna 1050}
\affiliation{Institute for High Energy Physics, Protvino 142281}
\affiliation{INFN - Sezione di Napoli, 80126 Napoli}
\affiliation{INFN - Sezione di Torino, 10125 Torino}
\affiliation{Advanced Science Research Center, Japan Atomic Energy Agency, Naka 319-1195}
\affiliation{J. Stefan Institute, 1000 Ljubljana}
\affiliation{Institut f\"ur Experimentelle Teilchenphysik, Karlsruher Institut f\"ur Technologie, 76131 Karlsruhe}
\affiliation{Kavli Institute for the Physics and Mathematics of the Universe (WPI), University of Tokyo, Kashiwa 277-8583}
\affiliation{Kennesaw State University, Kennesaw, Georgia 30144}
\affiliation{Department of Physics, Faculty of Science, King Abdulaziz University, Jeddah 21589}
\affiliation{Kitasato University, Sagamihara 252-0373}
\affiliation{Korea Institute of Science and Technology Information, Daejeon 34141}
\affiliation{Korea University, Seoul 02841}
\affiliation{Kyungpook National University, Daegu 41566}
\affiliation{P.N. Lebedev Physical Institute of the Russian Academy of Sciences, Moscow 119991}
\affiliation{Faculty of Mathematics and Physics, University of Ljubljana, 1000 Ljubljana}
\affiliation{Ludwig Maximilians University, 80539 Munich}
\affiliation{Luther College, Decorah, Iowa 52101}
\affiliation{Malaviya National Institute of Technology Jaipur, Jaipur 302017}
\affiliation{University of Maribor, 2000 Maribor}
\affiliation{Max-Planck-Institut f\"ur Physik, 80805 M\"unchen}
\affiliation{School of Physics, University of Melbourne, Victoria 3010}
\affiliation{University of Mississippi, University, Mississippi 38677}
\affiliation{University of Miyazaki, Miyazaki 889-2192}
\affiliation{Moscow Physical Engineering Institute, Moscow 115409}
\affiliation{Graduate School of Science, Nagoya University, Nagoya 464-8602}
\affiliation{Kobayashi-Maskawa Institute, Nagoya University, Nagoya 464-8602}
\affiliation{Universit\`{a} di Napoli Federico II, 80126 Napoli}
\affiliation{Nara Women's University, Nara 630-8506}
\affiliation{National Central University, Chung-li 32054}
\affiliation{National United University, Miao Li 36003}
\affiliation{Department of Physics, National Taiwan University, Taipei 10617}
\affiliation{H. Niewodniczanski Institute of Nuclear Physics, Krakow 31-342}
\affiliation{Nippon Dental University, Niigata 951-8580}
\affiliation{Niigata University, Niigata 950-2181}
\affiliation{Novosibirsk State University, Novosibirsk 630090}
\affiliation{Osaka City University, Osaka 558-8585}
\affiliation{Pacific Northwest National Laboratory, Richland, Washington 99352}
\affiliation{Panjab University, Chandigarh 160014}
\affiliation{Peking University, Beijing 100871}
\affiliation{University of Pittsburgh, Pittsburgh, Pennsylvania 15260}
\affiliation{Punjab Agricultural University, Ludhiana 141004}
\affiliation{Research Center for Nuclear Physics, Osaka University, Osaka 567-0047}
\affiliation{Department of Modern Physics and State Key Laboratory of Particle Detection and Electronics, University of Science and Technology of China, Hefei 230026}
\affiliation{Seoul National University, Seoul 08826}
\affiliation{Showa Pharmaceutical University, Tokyo 194-8543}
\affiliation{Soochow University, Suzhou 215006}
\affiliation{Soongsil University, Seoul 06978}
\affiliation{Sungkyunkwan University, Suwon 16419}
\affiliation{School of Physics, University of Sydney, New South Wales 2006}
\affiliation{Department of Physics, Faculty of Science, University of Tabuk, Tabuk 71451}
\affiliation{Tata Institute of Fundamental Research, Mumbai 400005}
\affiliation{Department of Physics, Technische Universit\"at M\"unchen, 85748 Garching}
\affiliation{School of Physics and Astronomy, Tel Aviv University, Tel Aviv 69978}
\affiliation{Toho University, Funabashi 274-8510}
\affiliation{Department of Physics, Tohoku University, Sendai 980-8578}
\affiliation{Earthquake Research Institute, University of Tokyo, Tokyo 113-0032}
\affiliation{Department of Physics, University of Tokyo, Tokyo 113-0033}
\affiliation{Tokyo Institute of Technology, Tokyo 152-8550}
\affiliation{Tokyo Metropolitan University, Tokyo 192-0397}
\affiliation{Utkal University, Bhubaneswar 751004}
\affiliation{Virginia Polytechnic Institute and State University, Blacksburg, Virginia 24061}
\affiliation{Wayne State University, Detroit, Michigan 48202}
\affiliation{Yamagata University, Yamagata 990-8560}
\affiliation{Yonsei University, Seoul 03722}
  \author{J.~Yelton}\affiliation{University of Florida, Gainesville, Florida 32611} 
  \author{I.~Adachi}\affiliation{High Energy Accelerator Research Organization (KEK), Tsukuba 305-0801}\affiliation{SOKENDAI (The Graduate University for Advanced Studies), Hayama 240-0193} 
  \author{J.~K.~Ahn}\affiliation{Korea University, Seoul 02841} 
  \author{H.~Aihara}\affiliation{Department of Physics, University of Tokyo, Tokyo 113-0033} 
  \author{S.~Al~Said}\affiliation{Department of Physics, Faculty of Science, University of Tabuk, Tabuk 71451}\affiliation{Department of Physics, Faculty of Science, King Abdulaziz University, Jeddah 21589} 
  \author{D.~M.~Asner}\affiliation{Brookhaven National Laboratory, Upton, New York 11973} 
  \author{T.~Aushev}\affiliation{Higher School of Economics (HSE), Moscow 101000} 
  \author{R.~Ayad}\affiliation{Department of Physics, Faculty of Science, University of Tabuk, Tabuk 71451} 
  \author{V.~Babu}\affiliation{Deutsches Elektronen--Synchrotron, 22607 Hamburg} 
  \author{S.~Bahinipati}\affiliation{Indian Institute of Technology Bhubaneswar, Satya Nagar 751007} 
% ...
\collaboration{The Belle Collaboration}


\begin{abstract}
Using 
980 ${\rm fb}^{-1}$ of data {collected} with the Belle detector
operating at the KEKB asymmetric-energy $e^+e^-$ collider, we {report a study} of the 
electromagnetic decays of 
excited {charmed baryons} $\Xi_c(2790)$ and 
$\Xi_c(2815)$. 
A clear signal (8.6 standard deviations) is  observed for 
$\Xi_c(2815)^0\to\Xi_c^0\gamma$, and we measure:
\begin{center}
\vspace{-0.1in}
$\mfrac {{\cal B}[\Xi_c(2815)^{0}\to\Xi_c^{0}\gamma] }{ {\cal B}[\Xi_c(2815)^0\to\Xi_c(2645)^+\pi^-\to\Xi_c^0\pi^+\pi^-]} = 0.41 \pm 0.05 \pm 0.03.$
\end{center}
We also present evidence (3.8 standard deviations) for the  
similar decay of the $\Xi_c(2790)^0$ and measure:
\begin{center}
$\mfrac{{\cal B}[\Xi_c(2790)^{0}\to\Xi_c^{0}\gamma]}{{\cal B}[\Xi_c(2790)^0\to\Xi_c^{\prime +}\pi^{-}\to\Xi_c^{+}\gamma \pi^-]} = 0.13 \pm 0.03 \pm 0.02$. 
\end{center}
The first quoted uncertainties are statistical and 
the second systematic.
We find no hint of the analogous decays of the $\Xi_c(2815)^+$ and $\Xi_c(2790)^+$ baryons and {set} upper limits
at the 90\% confidence level of:
$\mfrac{ {\cal B}[\Xi_c(2815)^{+}\to\Xi_c^{+}\gamma]}{{\cal B}[\Xi_c(2815)^+\to\Xi_c(2645)^0\pi^+\to\Xi_c^+\pi^-\pi^+]} < 0.09,$ 
and $\mfrac{ {\cal B}[\Xi_c(2790)^{+}\to\Xi_c^{+}\gamma]}{{\cal B}[\Xi_c(2790)^+\to\Xi_c^{\prime 0}\pi^{+}\to\Xi_c^{0}\gamma \pi^+]} < 0.06.$
Approximate values of the partial widths of the decays are extracted, 
{which} can be used to
discriminate between models of the underlying quark structure of these excited states.
\end{abstract}


\maketitle


{\renewcommand{\thefootnote}{\fnsymbol{footnote}}}
\setcounter{footnote}{0}


The $\Xi_c$ baryons comprise {$csu$ or $csd$ quark combinations~\cite{CC}.} 
Many excited states of these baryons
have been observed and studied~\cite{PDG}. 
In particular, a recent study~\cite{JMY} {reported} measurements of 
the masses and widths of the $\Xi_c(2790)^{+/0}$ and
$\Xi_c(2815)^{+/0}$ states. In the picture of a charmed baryon comprising a heavy ($c$) quark and a light ($su$ or $sd$) diquark,
these states are typically interpreted as $L=1$ orbital excitations of the ground {states where} the unit of angular momentum
is between the charm quark and a spin-0 light diquark system~\cite{Four1,Four2,Four3,Four4,Four5}. 
Such excitations are denoted $\lambda$ excitations. In this model, the $\Xi_c(2790)$ is
the $J^P=\frac{1}{2}^-$ state and the $\Xi_c(2815)$ the $J^P=\frac{3}{2}^-$ state, and the particles recently observed at
higher masses by LHCb~\cite{LHCb} are part of the expected family of corresponding states with a spin-1 diquark.  
These identifications are not made by direct
measurement of the spin and parity of the {states,} rather by inspection of {their} 
mass spectra and observed decay modes; clearly other 
interpretations are possible~\cite{Oset}.

In general, the decays of excited charmed baryons proceed via strong interactions, with the only electromagnetic decays
observed so far being $\Xi_c^{\prime}\to\Xi_c\gamma$~\cite{JESS,JMY} and 
$\Omega_c(2770)\to\Omega_c\gamma$~\cite{AUB,SOLOV}, 
since for these transitions the mass difference is not
sufficient for a strong decay. However, some predictions for the partial {widths} of photon transitions indicate that they could 
be observable. In particular, one theoretical treatment by Wang, Yao, Zhong, and Zhao (WYZZ)~\cite{theory} 
{predicts 
a partial width} of 263~${\rm keV/}c^2$
for the decay  $\Xi_c(2790)^0\to\Xi_c^0\gamma$ and 292~${\rm keV/}c^2$ for $\Xi_c(2815)^0\to\Xi_c^0\gamma$, 
assuming that they are $\lambda$ excitations. 
\textcolor{black}{On the other hand, the}
analogous decays for the $\Xi_c^+$
baryons are predicted to have very small partial widths. The same model predicts widths of $<10\ {\rm keV}/c^2$  
{if the unit of orbital excitation is}
between the two light quarks (a ``$\rho$ excitation''). Other models make different predictions~\cite{IKLR}; in particular, a treatment of
the $\Xi_c(2790)$ isodoublet as dynamically generated baryons predicts large partial widths for both charge states~\cite{GAME}. 
These predictions are summarized in Table~\ref{tab:Table1}.

{In this paper, we present}
 a search for the electromagnetic decays { $\Xi_c(2790,2815)^{+/0}\to\Xi_c^{+/0}\gamma$. The results} 
{are} converted to branching ratios and, with certain assumptions, to estimates of the partial widths for these
decays. 
These estimates can then be compared to the theoretical models and thus probe the inner structure of 
these {heavy} baryons.

\begin{table*}[htb]
\caption{Theoretical {predictions} of the partial widths in ${ \rm keV}/c^2$ for the $\Xi_c(2790)$ and $\Xi_c(2815)$. There are three
predictions from WYZZ~\cite{theory} as they model \textcolor{black}{one} $\lambda$ and two $\rho$ excitation states for each overall $J^P$.
The experimental measurements of the total widths are also listed.
}

\begin{tabular}
 { c | r | c | r| c |c |c }

\hline \hline
Mode & \multicolumn{3}{c|}{WYZZ~\cite{theory}}   & IKLR~\cite{IKLR} &GJR~\cite{GAME} &Actual total  \\
    &  $\lambda$ excitation & $\rho$ excitation&  $\rho$ excitation & $\lambda$ excitation&dynamically generated states & width~\cite{JMY} \
 \\

\hline
$\Xi_c(2790)^+\to\Xi_c^+\gamma$     &4.65\phantom{2}    &1.39 & 0.79\phantom{85} &  {--} & 246 & $8900\pm600\pm800$\\
$\Xi_c(2790)^0\to\Xi_c^0\gamma$     &263\phantom{.652}    &5.57& 3.00\phantom{85} &  -- & 117 & $10000\pm700\pm800$\phantom{2}\\
$\Xi_c(2815)^+\to\Xi_c^+\gamma$     &2.8\phantom{52}    &1.88 & 2.81\phantom{85} &  $190\pm\phantom{1}5$&-- &$2430\pm200\pm170$\\
$\Xi_c(2815)^0\to\Xi_c^0\gamma$     &292\phantom{.652}    &7.50 & 11.2\phantom{185} &  $497\pm 14$& -- &$2540\pm180\pm170$\\

\hline
\hline
\end{tabular}
\label{tab:Table1}
\end{table*}

The Belle detector~\cite{Belle} was a large-solid-angle spectrometer 
operating at the KEKB asymmetric-energy $e^+e^-$ collider~\cite{KEKB}, 
comprising 
six subdetectors: {the tracking system composed of the silicon vertex detector 
and the 50-layer central
drift chamber, the aerogel Cherenkov counter, the time-of-flight scintillation counter,} 
the electromagnetic calorimeter, and the 
$K_L^0$ and muon detector. A superconducting solenoid produced a 1.5 T magnetic field throughout the first five of these subdetectors.
Two inner detector configurations were used. 
{The first }
{consisted of} 
a 3-layer silicon vertex detector and a 2.0~cm radius beampipe, 
and the second of a 4-layer silicon detector and a small-cell inner drift chamber around a 1.5~cm radius beampipe.

\begin{figure}[htb]                                                                                                                   

\includegraphics[width=3.5in]{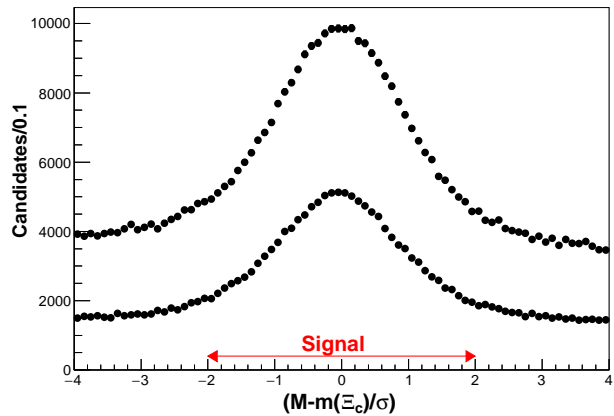}
\caption{Pull mass distribution for the $\Xi_c^0$ (upper data points), and $\Xi_c^+$ (lower data points) candidates.}

\label{fig:Figure1}

\end{figure}

In order to study $\Xi_c$ baryons, we first reconstruct a large sample of ground-state $\Xi_c^0$ and $\Xi_c^+$
baryons with good signal-to-noise ratio. To obtain large statistics, we use ten decay modes of the 
$\Xi_c^0$, and seven of the $\Xi_c^+$ ground states, as used in Ref.~\cite{JMY}. 
The decays are reconstructed from combinations of charged particles
measured using the tracking system, and neutral particles measured in the electromagnetic
calorimeter. The decays of long-lived mesons and hyperons are measured using secondary and tertiary vertex reconstruction. 
Each mode has specific requirements on its
decay products designed to suppress combinatorial backgrounds, and we follow the selection criteria described 
{in detail in our} 
previous publication~\cite{JMY}, 
{except
for the requirement} on the momentum of the $\Xi_c$ in the center-of-mass frame, $p^*,$ 
which is set as $p^* > 2.25\ {\rm GeV}/c$, 
{a choice which} is described below.
To show the yield of the reconstructed $\Xi_c^0$ and $\Xi_c^+$ baryons, 
we present in Fig.~\ref{fig:Figure1} 
the distributions of ``pull mass'',
i.e., the difference between the measured and nominal {mass} (2470.91 ${\rm MeV}/c^2$ and 2467.93
${\rm MeV}/c^2$ for the $\Xi_c^0$ and $\Xi_c^+$, respectively~\cite{PDG}), divided by the resolution 
($\sigma$), which is found mode-by-mode and is $\sim 5\ {\rm MeV}/c^2$.  
$\Xi_c$ candidates are selected if they are within $\pm 2\sigma$ of the nominal mass. For $\Xi_c^+$, the
number of selected candidates is 79k above a background of 61k, and for 
$\Xi_c^0$ 142k
signal candidates with a background of 154k.

To optimize the requirements specific to this analysis, a simulated data set \textcolor{black}{is} constructed using
the combination of the decays under study and generic $e^+e^-$ hadronic events. In addition to the 
$p^* >2.25\ {\rm GeV}/c$
requirement on the $\Xi_c$ momentum, the following three 
{selection criteria}
\textcolor{black}{are}
determined by maximizing the signal 
significance in 
{the} 
sample. {First, the} photon energy
is required to be greater than $550\ {\rm MeV}$. {Second, the} sum of the energy deposited in the central nine cells of 
a $5\times 5$ cell photon 
cluster is required to {be} at least 94\% of the total energy of the cluster.  
{Third, to discriminate against photons that are
$\pi^0$ daughters,} 
each photon is combined with each other photon candidate in the event and the pair 
{is
rejected if the likelihood of it being part of a $\pi^0$ is larger than 0.5.} These likelihoods are determined from Monte Carlo (MC) 
studies~\cite{Kopp} and are a function of the {energy} of the other photon, its polar \textcolor{black}{angle,} and the mass of the 
two-photon system. 
This last requirement retains  
{87\% of \textcolor{black}{the} signal 
according to Monte Carlo studies, while eliminating 42\% of the background.} 

\begin{figure}[htb]                                                                                                                   
\hspace{-.2in}
\includegraphics[width=4in]{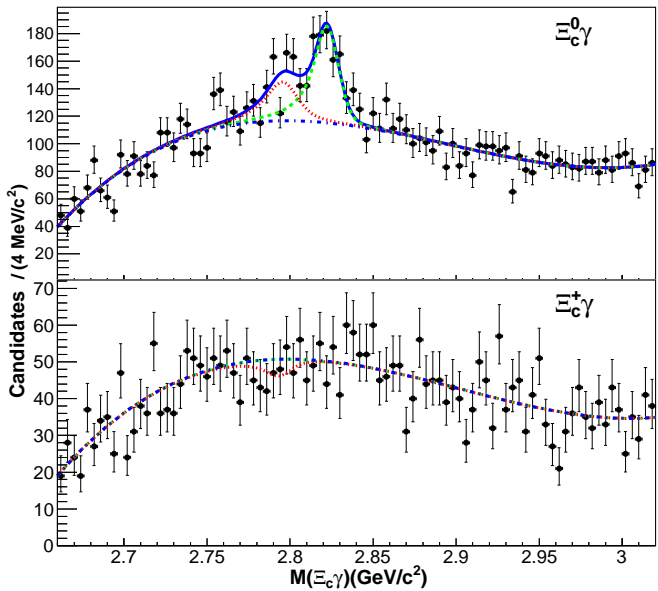}
\caption{The $\Xi_c\gamma$ mass distributions for (upper) $\Xi_c^0$ and (lower) $\Xi_c^+$. The fits are described in the text.
In addition to the total fitted yields, the fitted $\Xi_c(2815)$ signal components (dotted lines, green) and
$\Xi_c(2790)$ components (dashed lines, red) are shown stacked above the combinatorial background (dot-dashed lines, blue).}
\label{fig:signal}

\end{figure}
Figure~\ref{fig:signal} shows the $\Xi_c\gamma$ 
{invariant-mass}
distributions for the charged and neutral $\Xi_c$ baryons.
We fit a sum of a polynomial and two signal functions to the distributions using a 
binned maximum-likelihood fit with fine mass bins. 
In each case,
the signal is a Breit-Wigner function 
convolved with a ``Crystal Ball'' function~\cite{CB} to represent the 
detector {resolution.}
The parameters of the latter function are found with a GEANT-based MC simulation~\cite{Geant} to model the
response of the detector. The photon energies in the simulation are corrected to take into account the data-MC difference of resolution 
based on studies of mass resolution in the
decays $\pi^0\to\gamma\gamma$, $\eta\to\gamma\gamma$, and $D^{*0}\to D^0\gamma$~\cite{Umberto, Mizuk}.  
The masses and widths of the four particles under consideration have been precisely measured
in our previous {analysis~\cite{JMY}} 
and 
{are thus} 
fixed to the values reported. The width of the resolution functions are $\sim6.5\ {\rm MeV}/c^2$ with an 
estimated systematic uncertainty of $3\%$,
so in each distribution the two signal functions overlap. In each case a third-order polynomial is used to describe the 
{combinatorial}
background.
There is a clear signal for the decay $\Xi_c(2815)^0\to\Xi_c^0\gamma$ with $401\pm45$ events
and evidence for the decay $\Xi_c(2790)^0\to\Xi_c^0\gamma$ with $222\pm55$ events.
The statistical significance of each signal is calculated by 
excluding 
{the respective peak from the fit and finding} 
the change in the log-likelihood
($\Delta[\ln L]$).
{
The significance is expressed} in terms of standard {deviations, $n_{\sigma}$,} 
using the formula $n_{\sigma} = \sqrt{2\Delta[\ln L]}$. 
{For the decays
$\Xi_c(2815)^0\to\Xi_c^0\gamma$
and  $\Xi_c(2790)^0\to\Xi_c^0\gamma$
we find $n_{\sigma}~=~9.7$ and
4.0, respectively.} 
No signals are present in the $\Xi_c^+\gamma$ mass distribution, and the fit
yields are $0\pm25$ and $-32\pm31$ decays of $\Xi_c(2815)^+$ and $\Xi_c(2790)^+$ 
{baryons, respectively.}
{In order} to find upper-limit \textcolor{black}{signal yields} from these decays,
{we use a second-order polynomial as the background function,} 
as its 
{reduced $\chi^2$
is satisfactory, and this produces a more conservative limit.
We calculate the upper limits by integrating the likelihood functions obtained 
from the fits, and \textcolor{black}{then finding the yield values} for which the integrals 
contain 90\% of the total integral of \textcolor{black}{positive yields.} (That is, we set a
Bayesian upper limit using a uniform prior on the yield).
 We find 
90\% confidence level limits of 56 and 64 events for the decays of the $\Xi_c^+(2815)$ and $\Xi_c^+(2790)$, respectively.}

The masses and widths of the excited $\Xi_c$ states are very well known and their uncertainties have negligible 
effect on these yields. For the two significant signals,  
the largest systematic uncertainty is due to uncertainties in the 
background shape, 
evaluated by noting the change in the yield found when increasing the order of the Chebychev
polynomial used for the background function (5\%); decreasing the order of the polynomial 
produces an unsatisfactory fit result and so is not used.
Taking into account this systematic uncertainty, we find 
{the 
significances of the signals for 
$\Xi_c(2815)^0\to\Xi_c^0\gamma$ 
and
$\Xi_c^0(2790)\to\Xi^0\gamma$ to be 
$n_{\sigma}= 8.6$} { and 
3.8, respectively.}

\begin{figure}[htb]
\hspace{-.1in}
\includegraphics[width=1.74in]{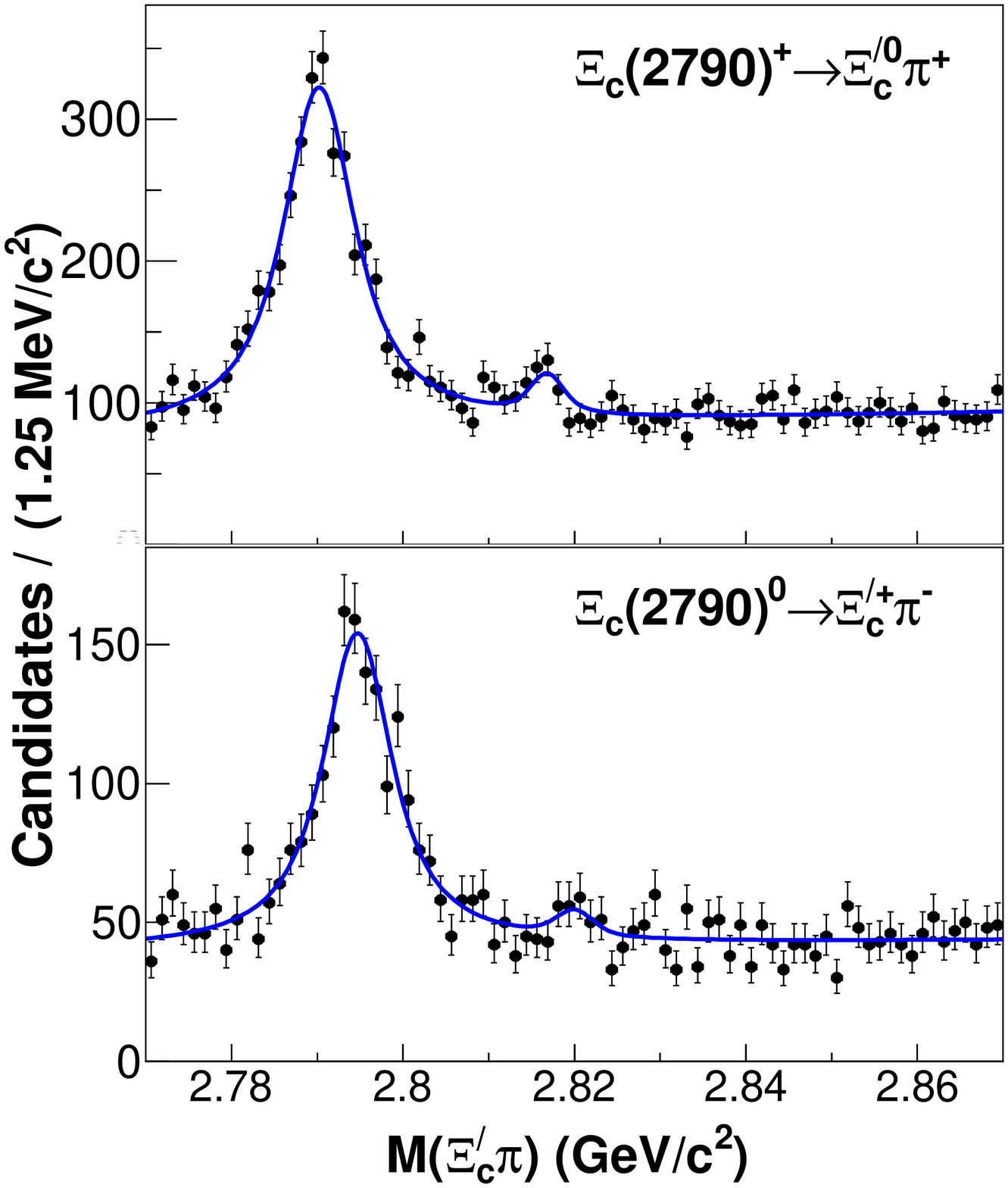}
\hspace{-.1in}
\includegraphics[width=1.74in]{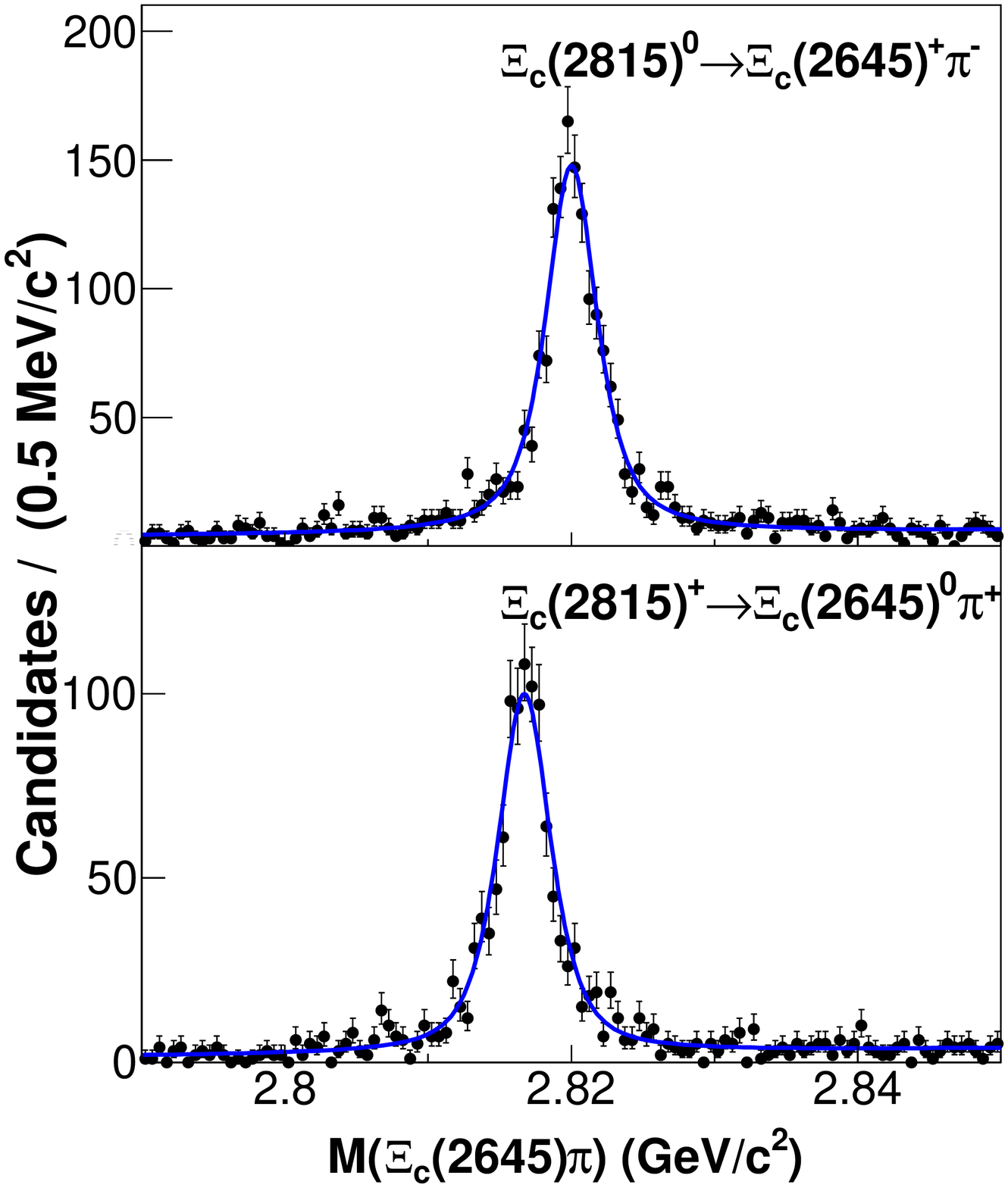}

\caption{The signals used as 
normalization modes in the analysis.
}
\label{fig:Figure3}

\end{figure}

To measure branching ratios
\newline
$R_{2815} = \mfrac{{\cal B}[\Xi_c(2815)^{+/0}\to\Xi_c^{+/0}\gamma]}{{\cal B}[\Xi_c(2815)^{+/0}\to\Xi_c(2645)^{0/+}\pi^{+/-}\to\Xi_c^{+/0}\pi^+\pi^-]}$
and 
$R_{2790} = \mfrac{{\cal B}[\Xi_c(2790)^{+/0}\to\Xi_c^{+/0}\gamma]}{{\cal B}[\Xi_c(2790)^{+/0}\to\Xi_c^{\prime 0/+}\pi^{+/-}\to\Xi_c^{0/+}\gamma\pi^{+/-}]}$,
\newline
we reconstruct the 
{normalization} 
modes following the technique presented in the previous Belle paper~\cite{JMY}, but using the 
momentum requirement on the daughter $\Xi_c$ baryons of $p^* > 2.25\ {\rm GeV}/c${.} 
The invariant-mass distributions for the normalization modes are shown in {Fig.~\ref{fig:Figure3},} and the yields for
the signals 
{listed} 
in Table~\ref{tab:yields}. 
For the measurement of {$R_{2815}$,} the largest systematic uncertainty is due to the signal-yield extraction of the 
electromagnetic decays
as detailed {above. In} addition, there are small 
{contributions due to}
the efficiency {estimation} of the photon {(3\%)~\cite{Umberto}},
uncertainties due to the modeling of the relative contributions of the different sub-modes (3\%),
the resolution of the $\Xi_c\gamma$ mass distribution (2\%), 
the uncertainty in the tracking efficiency (2\%), 
the fitting of the {normalization} mode (1\%), 
and 
{uncertainties due to the Monte Carlo statistics used to evaluate
efficiencies (1\%).} 
For the neutral {mode,} 
we \textcolor{black}{find} a value of $R_{2815} = 0.41\pm0.05\pm0.03$. For the charged mode, 
where no signal is observed, we set
{a} limit at 90\% confidence level of $R_{2815} < 0.09$.

The calculation of the $R_{2790}$ branching ratios has the {complication} 
that the signal and normalization modes involve decays 
into different ground-state charmed baryons. 
Our determination of the relative reconstruction efficiency of the $\Xi_c^0$
with respect to the $\Xi_c^+$ depends on the relative production rate of the two
states in the Belle dataset, which is not well known.
We make the assumption that the production of $\Xi_c^0$ and $\Xi_c^+$ with $p^*> 2.25\ {\rm GeV}/c$ 
{
is equal, which would be the case with exact isospin symmetry between the $u$ and $d$ quarks.
Deviations from this equality can occur if the probability of \textcolor{black}{creating} 
an $su$ or an $sd$ diquark in the fragmentation
process is different.
\textcolor{black}{In addition,} the decays from excited particles will not exactly preserve
isospin symmetry because of the isospin mass splitting of several ${\rm MeV}/c^2$ that has been measured in $\Xi_c$
ground states and some excited states~\cite{PDG}, and also is present in $\pi$ mesons. 
}
We estimate the systematic uncertainty associated with the equality
assumption to be $\pm 15\%$; this is larger than the asymmetry observed in the
$\Sigma_c^{++}/\Sigma_c^0$ system~\cite{SIGC}.

We find $R_{2790}$ = $0.13\pm0.03\pm0.02$ for the decay of the $\Xi_c(2790)^0$. For the decay of the $\Xi_c^+$
we {set} a limit at 90\% confidence level of $R_{2790} < 0.06$.

\begin{table}[h]

\caption{Yields of the normalization modes found from fits to the distributions shown in 
Fig.~\ref{fig:Figure3}. In all cases, there is a requirement on the momentum
of the ground-state charmed baryon of $p^* > 2.25\ {\rm GeV}/c${.}}

\begin{tabular}
 { c | c }

\hline \hline
Decay & Yield\\
\hline
$\Xi_c(2790)^+\to\Xi_c^{\prime 0}\pi^+\to\Xi_c^0\gamma\pi^+$ & $2591\pm140$\\
$\Xi_c(2790)^0\to\Xi_c^{\prime +}\pi^-\to\Xi_c^+\gamma\pi^-$ & $1231\pm\phantom{1}87$\\
$\Xi_c(2815)^0\to\Xi_c(2645)^+\pi^-\to\Xi_c^0\pi^+\pi^-$     & $1646\pm\phantom{1}50$\\
$\Xi_c(2815)^+\to\Xi_c(2645)^0\pi^+\to\Xi_c^+\pi^-\pi^+$     & $1121\pm\phantom{1}40$\\

\hline
\hline
\end{tabular}

\label{tab:yields}
\end{table}

We cannot directly measure the partial widths of the decay modes under consideration. However, we can use our branching ratio 
measurements, together with the already measured total widths~\cite{JMY}, to make estimates of the partial widths which can then be 
compared with theory. For the case of $\Xi_c(2815)\to\Xi_c(2645)\pi\to\Xi_c\pi\pi$ we calculate, using Clebsch-Gordan coefficients
and phase space, that the charged-pion decays account for $(38\pm4)\%$ of the total rate of this decay chain, where the rest of the
decays include $\pi^0$ transitions. The uncertainty in this number takes into account the mass and width uncertainties of the
excited states, and is an estimate as none of the $\pi^0$ transitions have been 
{observed and} isospin is not an exact symmetry.
Taking into account the decays $\Xi_c(2815)\to\Xi_c^{\prime}\pi$ measured previously~\cite{JMY}, 
the width of the electromagnetic decay is observed to be $(13.6\pm1.5 \pm1.7)\%$ of the total width, 
where the first uncertainty is statistical, and the second
is systematic.
There is an additional possibility that 
other decays exist that we do not detect. These include possible single-pion decays from the orbitally excited states to the 
ground state, double-pion decays that do not go through an intermediate resonance, and {transitions} 
that involve electromagnetic
decays to or from intermediate states. None of these are expected to be large, and we can estimate that they will produce a
reduction of the calculated partial width of no more than 20\%.
Based on these considerations, we estimate
a partial width of $\Gamma[\Xi_c(2815)^0\to\Xi_c^0\gamma] =  320 \pm 45 ^{+45}_{-80}\ {\rm keV}/c^2$. 
For the decays of the $\Xi_c(2815)^+$
we use similar arguments to find $\Gamma[\Xi_c(2815)^+\to\Xi_c^+\gamma] <  80~{\rm keV}/c^2$.

For the $\Xi_c(2790)^0$ we find that a similar calculation leads to $(7.9 \pm 2.0^{+1.7}_{-2.3}) \%$ 
of the total width being due to the 
electromagnetic decay, implying a partial width 
{
of $\Gamma[\Xi_c(2790)^0\to\Xi_c^0\gamma] \sim 800\ {\rm keV}/c^2$
with an uncertainty of around 40\%.}  
Similarly, for the decay $\Xi_c(2790)^+\to\Xi_c^+\gamma$, for which no signal is found, the upper limit on the partial width is set at
 $ 350\ {\rm keV}/c^2$. 

{The difference between the decays of the neutral and charged $\Xi_c(2815)$ states is clear}, and these 
results are in good agreement with the prediction that
was based on an identification of the $\Xi_c(2815)$ as $\lambda$ orbital excitations of the ground-state baryons~\cite{theory}. For the
$\Xi_c(2790)$ decays, the data are
 much less precise. Still, the evidence for the decay of the neutral $\Xi_c(2790)$ and the absence
of evidence for its isospin partner  is consistent with these predictions.

To conclude, we report the first observation of an electromagnetic decay of an orbitally-excited charmed baryon, and measure the
branching ratio 
$\mfrac{{\cal B}[\Xi_c(2815)^{0}\to\Xi_c^{0}\gamma]}{{\cal B}[\Xi_c(2815)^0\to\Xi_c(2645)^+\pi^-\to\Xi_c^0\pi^+\pi^-]} = 0.41 \pm 0.05 \pm 0.03.$
We also present evidence for the similar decay of the $\Xi_c^0(2790)$ and measure
{
$\mfrac{{\cal B}[\Xi_c(2790)^{0}\to\Xi_c^{0}\gamma]}{{\cal B}[\Xi_c(2790)^0\to\Xi_c^{\prime +}\pi^{-}\to\Xi_c^{+}\gamma \pi^-]} = 0.13 \pm 0.03 \pm 0.02$.
}
We find no evidence of the analogous decays of the $\Xi_c(2815)^+$ and $\Xi_c(2790)^+$ baryons. 
Using reasonable estimates of the unseen decays, we conclude that the partial widths of the electromagnetic decays of the 
$\Xi_c(2815)^0$ and $\Xi_c(2790)^0$ into the ground states are $320\pm 45 ^{+45}_{-80}\ {\rm keV}/c^2$ 
and $\sim800\ {\rm keV}/c^2${,}
respectively. The partial widths for the similar decays of the $\Xi_c(2815)^+$ and $\Xi_c(2790)^+$ are 
less than $80\ {\rm keV}/c^2$ 
and less than $350\ {\rm keV}/c^2 $, respectively. 
These results are consistent with predictions based on the identification of the $\Xi_c(2815)$ and $\Xi_c(2790)$ baryons 
as 
orbital excitations of the $\Xi_c$ baryons{,} where the unit of orbital excitation is between the heavy quark and the
spin-0 light diquark system.

We thank the KEKB group for the excellent operation of the
accelerator; the KEK cryogenics group for the efficient
operation of the solenoid; and the KEK computer group, and the Pacific Northwest National
Laboratory (PNNL) Environmental Molecular Sciences Laboratory (EMSL)
computing group for strong computing support; and the National
Institute of Informatics, and Science Information NETwork 5 (SINET5) for
valuable network support.  We acknowledge support from
the Ministry of Education, Culture, Sports, Science, and
Technology (MEXT) of Japan, the Japan Society for the 
Promotion of Science (JSPS), and the Tau-Lepton Physics 
Research Center of Nagoya University; 
the Australian Research Council including grants
DP180102629, 
DP170102389, 
DP170102204, 
DP150103061, 
FT130100303; 
Austrian Science Fund (FWF);
the National Natural Science Foundation of China under Contracts
No.~11435013,  
No.~11475187,  
No.~11521505,  
No.~11575017,  
No.~11675166,  
No.~11705209;  
Key Research Program of Frontier Sciences, Chinese Academy of Sciences (CAS), Grant No.~QYZDJ-SSW-SLH011; 
the  CAS Center for Excellence in Particle Physics (CCEPP); 
the Shanghai Pujiang Program under Grant No.~18PJ1401000;  
the Ministry of Education, Youth and Sports of the Czech
Republic under Contract No.~LTT17020;
the Carl Zeiss Foundation, the Deutsche Forschungsgemeinschaft, the
Excellence Cluster Universe, and the VolkswagenStiftung;
the Department of Science and Technology of India; 
the Istituto Nazionale di Fisica Nucleare of Italy; 
National Research Foundation (NRF) of Korea Grant
Nos.~2016R1\-D1A1B\-01010135, 2016R1\-D1A1B\-02012900, 2018R1\-A2B\-3003643,
2018R1\-A6A1A\-06024970, 2018R1\-D1A1B\-07047294, 2019K1\-A3A7A\-09033840,
2019R1\-I1A3A\-01058933;
Radiation Science Research Institute, Foreign Large-size Research Facility Application Supporting project, the Global Science Experimental Data Hub Center of the Korea Institute of Science and Technology Information and KREONET/GLORIAD;
the Polish Ministry of Science and Higher Education and 
the National Science Center;
the Ministry of Science and Higher Education of the Russian Federation, Agreement 14.W03.31.0026; 
University of Tabuk research grants
S-1440-0321, S-0256-1438, and S-0280-1439 (Saudi Arabia);
the Slovenian Research Agency;
Ikerbasque, Basque Foundation for Science, Spain;
the Swiss National Science Foundation; 
the Ministry of Education and the Ministry of Science and Technology of Taiwan;
and the United States Department of Energy and the National Science Foundation.


\begin{thebibliography}{99}
\bibitem{CC}
Throughout this paper, 
the inclusion of the charge-conjugate {is implied.}
\end{thebibliography}
\end{document}